\documentclass{article}
\usepackage[a4paper,margin=1in]{geometry}
\usepackage{graphicx}
\usepackage{hyperref}
\usepackage{comment}
\usepackage{authblk}
\usepackage[toc,page]{appendix}
\usepackage{amsfonts}
\title{A Comprehensive Review of AI Agents: Transforming Possibilities in Technology and Beyond}
\date{} 
\author[1]{Xiaodong Qu}
\author[2]{Andrews Damoah}
\author[3]{Joshua Sherwood}
\author[4]{Peiyan Liu}
\author[5]{Christian Shun Jin}
\author[6]{Lulu Chen}
\author[7]{Minjie Shen}
\author[8]{Nawwaf Aleisa}
\author[9]{Zeyuan Hou}
\author[10]{Chenyu Zhang}
\author[11]{Lifu Gao}
\author[12]{Yanshu Li}
\author[13]{Qikai Yang}
\author[14]{Qun Wang}
\author[15]{Cristabelle De Souza}

\affil[1,3,4,8,10,11]{George Washington University}
\affil[2]{University of Maryland, College Park}
\affil[5]{Independent Researcher}
\affil[6,7,9]{Virginia Tech}
\affil[12]{Brown University}
\affil[13]{University of Illinois Urbana-Champaign}
\affil[14]{San Francisco State University}
\affil[15]{Stanford University}

\begin{document}

\maketitle

\begin{abstract}
Artificial Intelligence (AI) agents have rapidly evolved from specialized, rule-based programs to versatile, learning-driven autonomous systems capable of perception, reasoning, and action in complex environments. The explosion of data, advances in deep learning, reinforcement learning, and multi-agent coordination have accelerated this transformation. Yet, designing and deploying unified AI agents that seamlessly integrate cognition, planning, and interaction remains a grand challenge. In this review, we systematically examine the architectural principles, foundational components, and emergent paradigms that define the landscape of contemporary AI agents. We synthesize insights from cognitive science-inspired models, hierarchical reinforcement learning frameworks, and large language model-based reasoning. Moreover, we discuss the pressing ethical, safety, and interpretability concerns associated with deploying these agents in real-world scenarios. By highlighting major breakthroughs, persistent challenges, and promising research directions, this review aims to guide the next generation of AI agent systems toward more robust, adaptable, and trustworthy autonomous intelligence.
\end{abstract}

\section{Introduction}

The development of artificial intelligence (AI) agents---autonomous systems capable of perceiving their surroundings, reasoning about possible courses of action, and executing decisions---has evolved significantly in recent decades. Early AI agents, rooted in the symbolic reasoning systems of the 1950s and 1960s, relied on hand-crafted rules and logic-based methods, excelling in constrained domains but struggling with adaptability and uncertainty\cite{stone2000multiagent, dosovitskiy2021vit}. The introduction of statistical learning and probabilistic reasoning in the 1980s and 1990s enhanced reliability, while the rise of reinforcement learning (RL) allowed agents to learn policies through trial-and-error interactions~\cite{feigenbaum1983fifth,ghallab2004automated, kaelbling1996reinforcement,sutton2018reinforcement}. The integration of deep neural networks with RL (DeepRL) led to breakthroughs such as superhuman performance in Atari games and Go~\cite{silver2016mastering,silver2017mastering}. With growing computational power, recent advancements in statistical methods and machine learning, AI agents have incorporated advanced perception, natural language sequence modeling, and cognitive-inspired principles, enabling them to adapt, collaborate, and mirror aspects of human reasoning in dynamic environments~\cite{dosovitskiy2021vit,devlin2019bert,krizhevsky2017imagenet,he2016resnet,brown2020language,lake2017building,marcus2020nextdecade}. 

Contemporary AI agents are increasingly deployed in high-stakes, real-world contexts: self-driving cars navigating congested urban environments~\cite{tang2024textsquare,paden2016survey}, autonomous laboratories accelerating scientific discovery~\cite{kitson2017configurable,zhao2024harmonizing}, virtual assistants managing complex user queries~\cite{tur2011spoken}, and automated trading agents operating in financial markets~\cite{deng2017deep}. Underpinning these successes are developments in deep learning for perception~\cite{krizhevsky2017imagenet,he2016resnet,dosovitskiy2021vit}, reinforcement learning (RL) for decision-making~\cite{mnih2015human,mnih2013playing}, large language models (LLMs) for communication and reasoning~\cite{devlin2019bert,brown2020language}, and multi-agent frameworks that orchestrate coordination and competition among numerous entities~\cite{silver2017mastering,yang2018mean}.

However, forging truly unified AI agents presents an array of open problems. Such agents must integrate perception, abstract reasoning, hierarchical planning, and flexible communication while ensuring safety, interpretability, and adherence to ethical standards. In this systematic review, we synthesize a broad literature on AI agents, examining foundational methods, current paradigms, and emerging architectures~\cite{zhao2024acellular}. We highlight key breakthroughs from cognitive-inspired models to LLM-driven reasoning engines, from hierarchical RL to multimodal sensor fusion, and from single-agent solutions to scalable multi-agent frameworks~\cite{wu2022cosbin}.

We also identify critical gaps and challenges. Robustness under domain shift, explainability of complex decision-making, and value alignment with human norms remain unsolved. Achieving human-level adaptability, transparency, resource efficiency, and trustworthy autonomy calls for deeper interdisciplinary research, from cognitive science to ethics, neuroscience, and economics.

\section{Related Work}
With the recent popularity of AI agents in both research and industry, alongside rapid advancements within this domain, we have begun to see an emergence of literature reviews attempting to consolidate and analyze the history and progress made within this evolving field.


\subsection{Past Reviews on AI Agents}
The review by Wang et al. takes a holistic approach when discussing the evolution of LLM-based agents~\cite{wang2024survey}. The authors focus on analyzing three main aspects including the foundations of AI agent construction, applications within various fields, and common evaluation strategies for benchmarking performance. They cover many of the core components and technology fueling AI agents, proposing a unified framework encompassing early development strategies, as well as the diverse applications of these agents in fields such as social science, natural science, and engineering. They also look at various strategies for evaluating AI agent performance, ranging from subjective evaluation, such as human annotations and Turing tests, to more objective metrics such as success rate and accuracy. Their review provides a comprehensive look into the development of AI agents and their domain applications within the current day.

The paper by Guo et al. conducts a similar review, focusing more on multi-agent systems, specifically for simulation research ~\cite{guo2024multiagent}. The study takes an in-depth look at the fundamental aspects of multi-agent systems, common domains that utilize multi-agent systems for simulations, and the challenges within this field. The authors compare differences in functionality between single-agent and multi-agent systems, specifically in the context of profiling, communications, and decision-making. They highlight the two main applications of multi-agent systems: problem-solving, which leverages collaboration to address complex tasks, and world simulation, where agents are used to replicate social environments to reproduce real-world interactions. The authors also address many of the common challenges within the field such as hallucinations, scaling, and lack of multi-modal support~\cite{zhu2025fdnet}.

The study conducted by Xi et al. investigates the role LLMs play as foundational models for AI agents~\cite{xi2023rise}. Through this lens, the authors detail the origins of LLM-based agents, applications in agent-to-agent (both human and artificial) interactions, and open questions in the field. They highlight key properties that validate LLMs as suitable foundations for agentic systems, such as autonomy and adaptability. They also highlight the importance of natural language processing, a key aspect of LLMs, in an agent's ability to reason and communicate. The authors discuss the deployment of AI agents in various scenarios including single-agent, multi-agent, and human-agent settings, as well as practical applications of each. They extend this methodology to agentic societies, where multiple agents interact to create an artificial society. This framework builds upon individual agent behaviors and environments to construct fully realistic, artificial societies.

The work detailed in the paper by Xie et al. focuses on a review of LLM-based agents specifically in the multimodal domain~\cite{xie2024multimodal}. They highlight the impact and challenges of multimodality on design frameworks, evaluation methods, and applications. The authors discuss how the core components of an AI agent, such as the planning and memory modules, require integration of textual, image, and audio capabilities into their architectures in multimodal settings. They also discuss the need for robust evaluation methods, especially for multimodal agents, that are capable of assessing the agent's capabilities across multiple domains and how they interact to allow for complex reasons and problem-solving. The authors also discuss the extensive and diverse applications of multimodal agents in fields like autonomous driving, game development, and robotics. They emphasize the importance of multimodality in driving human-computer interaction between humans and autonomous agents by enabling their use in more complex, nuanced settings.

As a collective, past reviews on AI agents develop a comprehensive assessment of the technology and provide an in-depth analysis of foundational elements, diverse applications, current evaluation paradigms, and future 
prospects of autonomous agents. Although many of these studies are suitable for people with prior knowledge in the field, few provide a simplified framework to understand these concepts, specifically targeting new researchers.







\section{Methodology}

We conducted a systematic literature review following the Preferred Reporting Items for Systematic Reviews and Meta-Analyses (PRISMA) guidelines to ensure a thorough and transparent approach. This method helped identify, screen, and select relevant studies focused on AI agents.\\

\begin{figure}[htp]
    \centering
    \includegraphics[width=10cm]{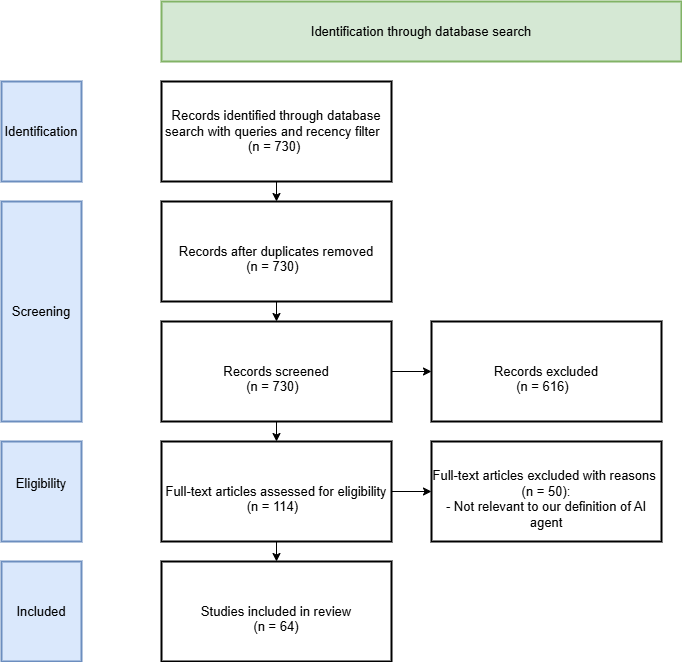}
    \caption{The PRISMA review was conducted independently by authors on each topic relating to core components, applications, and paradigm-shifting designs within the AI agent space, and combined into one diagram.}
    \label{fig:memory}
\end{figure}


\subsection{Search Strategy}

Following the PRISMA guidelines, we conducted our literature review process to identify relevant papers on foundational frameworks, current trends, and breakthrough advancements in AI agent research. Employing this strategy allowed us to conduct a rigorous analysis of relevant research, aligning with methodologies utilized in previous works. We used Google Scholar to conduct our review of emergent technologies in the AI agent space and 4 application domains: business, education, science, and entertainment. All queries were conducted using search terms related to AI agents and the individual topics. Common search terms included: "Autonomous Agent", "Reinforcement Learning Agent", and "Multi-Agent System", with "AI Agent" being the prime term used across all contributors' searches. The full list of queries by topic can be found in Appendix A.\\

\noindent The review process included three stages:
\begin{enumerate}
    \item \textbf{Google Scholar Search}: The records resulting from the query were examined by page, and full pages were included in order of relevancy (by Google Scholar's algorithm) until no results on the given page were relevant, based on titles.
    \item \textbf{Initial Screening}: Titles and abstracts were reviewed to exclude irrelevant papers.
    \item \textbf{Full-Text Review}: Full papers were reviewed to exclude irrelevant papers.
\end{enumerate}


%



\subsection{Exclusion Criteria}


To ensure the relevance and accessibility of our selected studies, we applied a rigorous set of exclusion criteria. One of the primary considerations was the availability of the full text. Studies that were only accessible as abstracts, summaries, or paywalled content without institutional or open-access availability were excluded. This step was necessary to ensure that all reviewed studies could be thoroughly analyzed, preventing misinterpretations or incomplete evaluations based on limited information~\cite{lu2021cot}.

Additionally, we filtered out publications that were not directly focused on AI agents or fell outside the intended scope of our review. Given the broad application of artificial intelligence across various domains, many papers may reference AI agents tangentially without providing substantial insights into their architectures, methodologies, or applications. To maintain a focused and cohesive analysis, we prioritized studies that directly contributed to AI agent research, whether in foundational theory, implementation strategies, or domain-specific applications.

Language accessibility was another key factor in our selection process. Since our review was conducted in English, we excluded papers published in other languages due to the potential for misinterpretation and difficulty in verifying content accuracy. While we acknowledge that valuable research exists in multiple languages, our methodology required consistency in analysis and comprehension, which was best achieved by limiting our review to English-language publications.

By implementing these exclusion criteria, we refined our dataset to ensure a high-quality and relevant selection of literature. Each filtered study contributed meaningfully to our analysis, helping to create a structured and comprehensive review of AI agents while avoiding redundancy and extraneous content~\cite{10.1145/3696271.3696292}.

\subsection{Data Extraction}
To systematically analyze the selected studies, we conducted a structured data extraction process, ensuring that each paper contributed valuable insights into AI agent research. The primary focus of this extraction was identifying the core methodologies used in AI agent development. This included examining the underlying frameworks and architectures employed in various studies, as well as the algorithms and techniques utilized for reasoning, decision-making, and adaptation. Understanding these methodological foundations allowed us to categorize and compare different approaches in AI agent design.

In addition to methodologies, we documented each study’s key findings and contributions to the field. This involved assessing the significance of their results, novel approaches, and the broader implications for AI agent research. Whether a study introduced a new decision-making framework, improved reinforcement learning strategies, or proposed a novel integration of multi-agent coordination, we aimed to capture its unique contribution to advancing AI intelligence and autonomy~\cite{wang2024application}.

Furthermore, we evaluated the relevance of each study to the progression of AI agent research. Studies that addressed pressing challenges, introduced paradigm-shifting ideas, or explored new application areas were given particular attention. By focusing on research that actively pushed the boundaries of AI agent capabilities, we ensured that our review highlighted the most impactful and forward-thinking work in the field.

\section{Results}

\subsection{Architectures and Learning Paradigms}

Modern AI agent architectures integrate diverse components, from memory systems to decision-making frameworks. For instance, \textit{Lilian Weng’s blog} provides a detailed exploration of agent architectures, emphasizing the interplay between memory, planning, and tool use. The \textit{Stanford HAI overview} highlights how computational agents are beginning to exhibit human-like behaviors, underscoring their potential for real-world applications. These studies collectively illustrate the shift toward more holistic and adaptive agent designs.

\subsection{Core Components of Modern AI Agent Architectures}

AI agent architectures include components designed to enable perception, reasoning, decision-making, and interaction. Our review provides a deep dive of these core components, highlighting key breakthroughs and future opportunities~\cite{wang2024self}.

The system contains four major components - memory, tools, actions and planning. Memory is divided into short-term and long-term memories, offering both contextual and enduring information. The Agent is also capable of employing a wide range of tools, such as calendars, calculators, code interpreters, and search functions to execute specialized tasks~\cite{wang2025fine}. Planning encompasses sophisticated techniques such as reflection, self-evaluation, chain of thought reasoning, and sub-goal breakdown, enabling the Agent to enhance its decision-making and tackle intricate challenges effectively.
(see Fig.~\ref{fig:overview}):
\begin{figure}[htp]
    \centering
    \includegraphics[width=15cm]{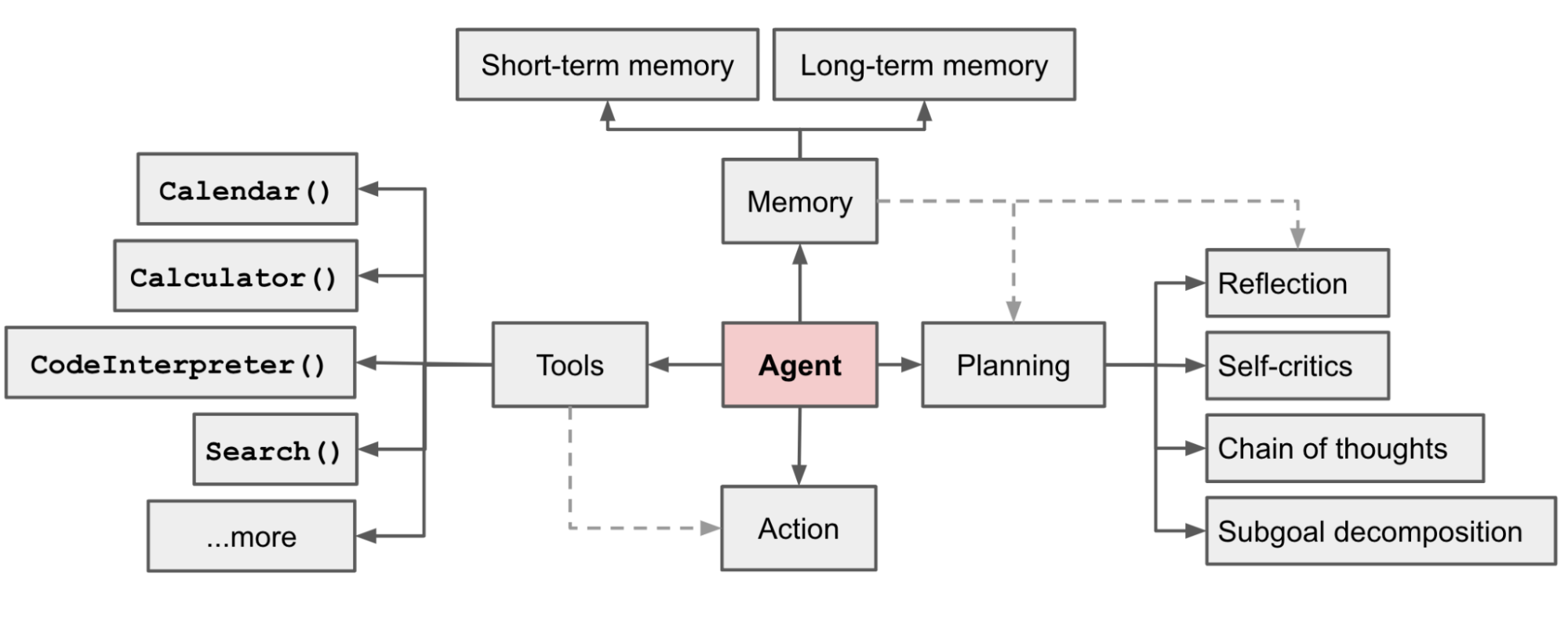}
    \caption{An overview of an AI Agent's core components ~\cite{weng2023agent}}
    \label{fig:overview}
\end{figure}

\subsection{Planning}
Inspired by classical AI planning~\cite{ghallab2004automated}, modern agents integrate symbolic and subsymbolic methods to reason about future states, causal dependencies, and long-term goals. Hierarchical RL structures decision-making at multiple levels of abstraction, improving sample efficiency and interpretability~\cite{kulkarni2016hierarchical,chitnis2021learning}. Additionally, model-based RL, graph-based planning, and hybrid neural-symbolic reasoning approaches allow agents to perform sophisticated problem-solving.

Chain-of-thought ~.\cite{wei2022chain} is a milestone technique to significantly improve the ability of large language models (LLMs) to perform complex reasoning tasks. The framework provides the models with examples of step-by-step reasoning (a "chain of thought") in the prompt, guiding them to break down complex problems into intermediate steps before arriving at the final answer. This helps the model perform better on tasks requiring multi-step reasoning. 

Agents must adapt policies over time. Reinforcement learning algorithms, from value-based methods~\cite{mnih2015human} to policy gradient techniques~\cite{schulman2017ppo,haarnoja2018sac}, enable agents to learn through interaction. 

Reflexion~\cite{shinn2024reflexion} is a framework designed to enhance AI Agents through reinforcement learning. It has a self-reflection architecture that leverages the heuristic function and linguistic feedback to improve reasoning skills. Reflexion represents a significant step forward in enabling LLM-based agents to learn from trial and error through verbal self-reflection. The framework’s ability to improve performance across diverse tasks without requiring extensive fine-tuning makes it a promising approach for future research in autonomous agents.

\begin{figure}[htp]
    \centering
    \includegraphics[width=15cm]{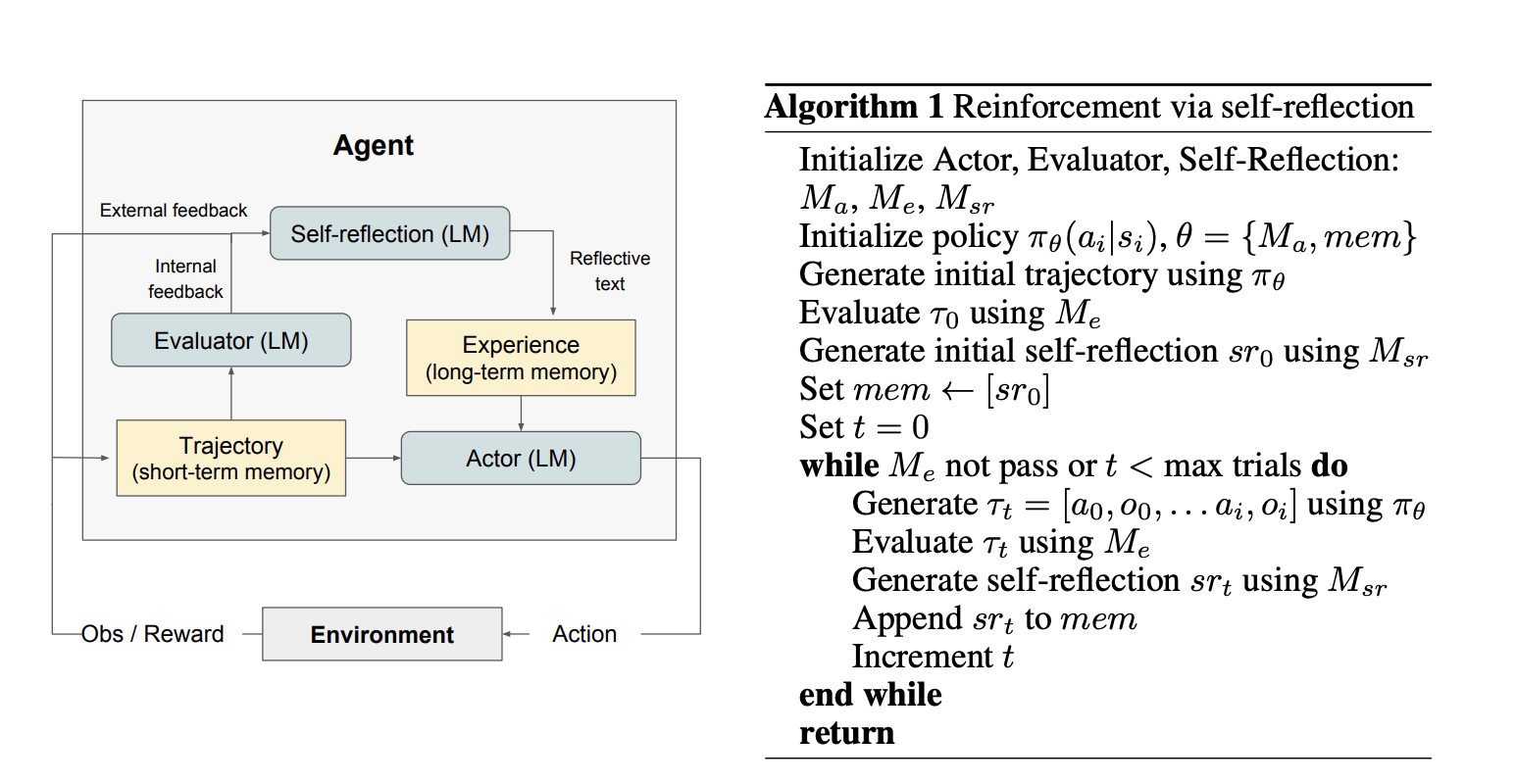}
    \caption{An overview of the Reflexion framework ~\cite{shinn2024reflexion}}
    \label{fig:relexion}
\end{figure}

Chain of Hindsight(CoH)~\cite{liu2023chain} is another framework that helps LLM agents improve their outputs by training with historical datasets that contain past sequential outputs with feedback. The framework aligns large language models with human feedback more effectively than traditional methods such as supervised fine-tuning (SFT) or reinforcement learning with human feedback (RLHF). The CoH algorithm trains the model to maximize the likelihood of predicting tokens in sequences $x = [x_1, x_2, \ldots, x_n]$, using a causal Transformer architecture:
\[
log p(x) = \sum_{i=1}^n \log p(x_i \mid x_{<i})
\]
The training sequence contains model outputs combined with feedback. To make sure the model will only be trained to predict non-feedback tokens, a masking technique is adopted such that:
\[
log p(x) = \sum_{i=1}^n \mathbb{1}_{O(x)}(x_i) \log p(x_i \mid x_{<i})
\]
where $\mathbb{1}_{O(x)}(x_i)$ is 1 if $x_i$ is not part of the feedback and 0 otherwise. Comprehensive experiments on summarization and dialogue datasets demonstrate that CoH significantly surpasses RLHF and other baseline methods.

Meta-learning~\cite{finn2017model} and continual learning~\cite{de2019continual} approaches allow agents to generalize across tasks and accumulate knowledge without catastrophic forgetting.

\subsection{Memory}
Memory processing in AI agents is crucial for enhancing their effectiveness. Much like human memory, it enables the acquisition, storage, retention, and retrieval of information for future use~\cite{weng2023agent}. Memory is categorized into short-term or textual, constrained by the context window of the underlying transformer models, and long-term or parametric, encompassing declarative (facts and events) and procedural (unconscious skills) memory~\cite{jiang2024ai}.

\begin{figure}[htp]
    \centering
    \includegraphics[width=15cm]{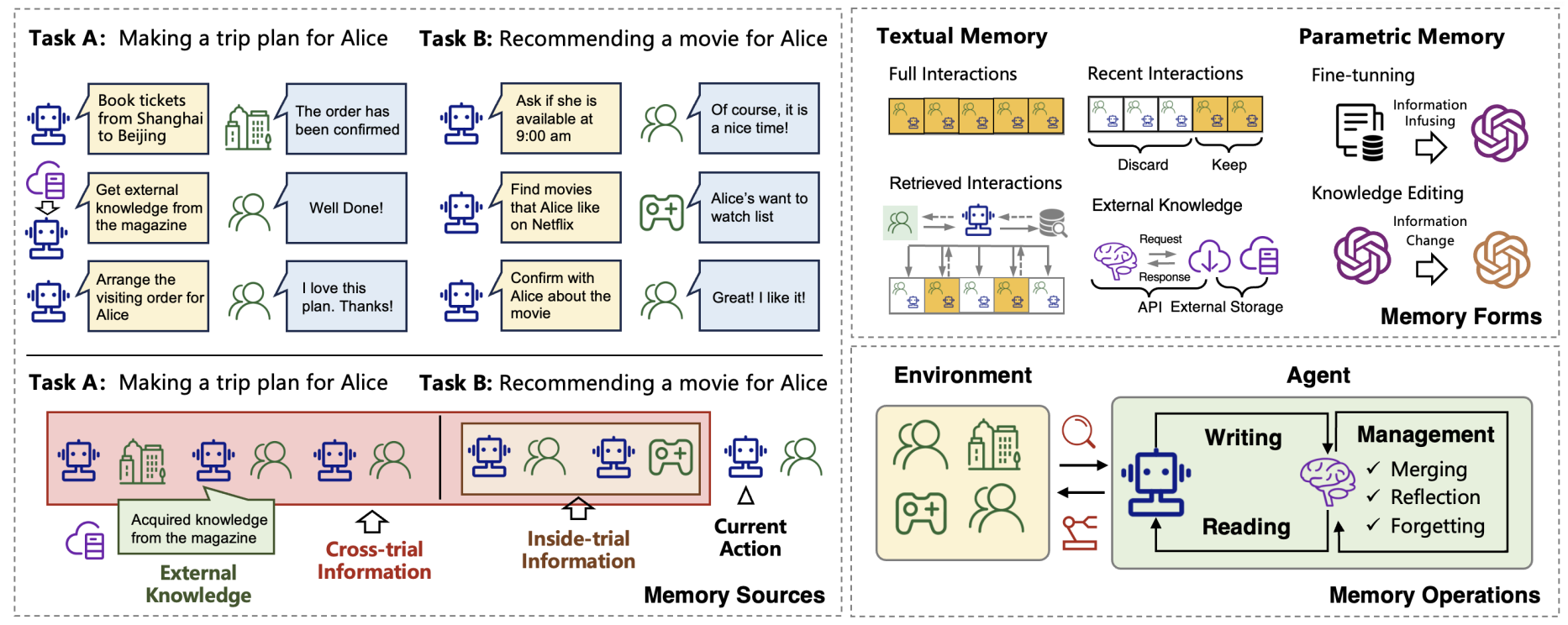}
    \caption{ An overview of the sources, forms, and operations of the memory in LLM-based agents. ~\cite{zhang2024survey}}
    \label{fig:memory}
\end{figure}

Memory-enhanced agents have demonstrated significant capabilities in maintaining context, emulating human-like behavior, and tackling complex tasks by effectively utilizing both short-term and long-term memory. However, despite these advancements, challenges remain in scalability, efficiency, and the seamless integration of external knowledge for LLM-based AI agents~\cite{su2024conflict}. Addressing these limitations is crucial for advancing LLM-based agents toward artificial general intelligence (AGI), enhancing their robustness and efficiency in real-world applications~\cite{xie2025theory,zhang2024survey}.

\subsection{Tools}
One of the hallmarks of human intelligence is the ability to use tools that reflect advanced cognitive functions such as problem-solving, adaptability, and foresight. Equipping LLMs with external tools can significantly extend model capabilities~\cite{weng2023agent}. In fact, LLM-based AI agents can be defined as systems where LLMs dynamically direct their own processes and tool usage, maintaining control over how they perform tasks~\cite{buildingagents}. Tools are external modules that LLMs can call to gather information or perform actions. Examples include information retrieval systems, search engines, code interpreters, and robotic arms~\cite{mialon2023augmented}. The key challenges for tool-augmented LLMs, as highlighted in the paper, revolve around hallucinations, planning complexity, and input errors, which greatly impact their viability in practical settings. These challenges underscore the necessity for more rigorous adherence to API documentation, enhanced planning algorithms, and broader, more inclusive training datasets to improve the robustness and reliability of AI agents in practical, real-world applications involving tool use~\cite{10.1145/3655497.3655500,li2023api}. 

\subsection{Perception Modules}
Perception converts raw sensory data (e.g., images, audio, text, LiDAR scans) into structured representations. Vision backbones rely on convolutional neural networks~\cite{krizhevsky2017imagenet,he2016resnet} and vision transformers~\cite{dosovitskiy2021vit}, while speech recognition and language processing leverage transformer-based LLMs~\cite{tang2022few,devlin2019bert,brown2020language}. Sensor fusion merges multiple modalities to form a holistic, environmental understanding.

\subsection{Representation and Abstraction Layers}
Agents require compact, meaningful representations. Self-supervised and contrastive learning approaches~\cite{chen2020simple,tang2022optimal} extract latent features from high-dimensional data. LLMs construct rich language embeddings that capture semantic and syntactic nuances, enabling agents to interpret instructions, query knowledge, and generate natural language outputs~\cite{10692439}.

\subsection{Planning and Reasoning Engines}
Inspired by classical AI planning~\cite{ghallab2004automated,tu2025active}, modern agents integrate symbolic and subsymbolic methods to reason about future states, causal dependencies, and long-term goals. Hierarchical RL structures decision-making at multiple levels of abstraction, improving sample efficiency and interpretability~\cite{kulkarni2016hierarchical,chitnis2021learning}. Additionally, model-based RL, graph-based planning, and hybrid neural-symbolic reasoning approaches allow agents to perform sophisticated problem-solving.

\subsection{Interaction and Communication Interfaces}
For agents to collaborate effectively, they must communicate. Natural language interfaces~\cite{devlin2019bert,brown2020language}, emergent communication protocols in MAS~\cite{lazaridou2019emergent}, grounded language acquisition~\cite{bisk2020experience}, and embodiment enable richer, more flexible multi-agent interactions and human-agent cooperation.


\section{Applications of AI Agents}
AI agents have begun to influence a broad spectrum of real-world applications, serving as critical components in complex decision-making processes and enhancing human capabilities. They integrate advances in perception, reasoning, communication, and control to provide adaptive and context-sensitive solutions that were once out of reach for traditional software systems~\cite{li2024advances, wu2025individual}.


\subsection{Healthcare}
AI agents—digital systems capable of perceiving their environment, reasoning about what they perceive, and taking actions - have begun to make a remarkable impact in healthcare~\cite{wang2025iddn}. As medical practices face growing patient demands, clinician burnout, and data complexity, these agents promise to lighten workloads, improve care quality, and augment the clinical decision-making process. While the concept of AI in medicine has existed for decades, recent advancements in computational power, machine learning algorithms, and the availability of large, high-quality datasets have accelerated progress, transforming what was once science fiction into everyday reality.
The development of AI agents in healthcare did not happen overnight. The rise of machine learning—particularly deep learning—over the past decade has dramatically changed the landscape. With the proliferation of electronic health records (EHRs), medical imaging databases~\cite{zhao2024two}, and wearable sensors, healthcare organizations today have unprecedented access to vast and varied datasets. This data revolution, combined with cheaper and more powerful cloud computing services, has allowed modern AI agents to become more accurate, context-aware, and integrated into existing healthcare workflows~\cite{key2024advancing,chen2021data}. By the late 2010s, AI agents began to show promise in detecting diseases from imaging scans, triaging patient queries, and even assisting in surgical interventions.

\subsubsection{Diagnostic and Decision-Support Agents}
Diagnostic and Decision-Support Agents
AI-driven clinical decision support systems (CDSS) harness large repositories of medical knowledge and patient data to guide clinicians~\cite{katz2024multi}. For example, agents trained on vast numbers of radiology images can highlight suspicious regions in a chest X-ray or mammogram, alerting a physician to potential conditions~\cite{li2024enhancing}. In dermatology, AI agents analyzing skin lesion images can help detect melanoma, while in ophthalmology, they can identify signs of diabetic retinopathy. These agents reduce the cognitive load on physicians, enhance diagnostic accuracy~\cite{10824316}, and potentially catch conditions earlier, leading to better patient outcomes.

\subsubsection{Patient-Facing Virtual Assistants and Chatbots}
On the patient side, AI-based virtual assistants and chatbots are becoming increasingly common. These conversational agents help patients book appointments, remind them to take medication, provide educational information, and even guide them through symptom checking~\cite{du2023abds}. By offering round-the-clock support, such agents improve access to care and empower patients to take more control of their health. They are especially helpful in primary care settings, where they can quickly triage simple queries and refer complex questions to healthcare professionals.

\subsubsection{Robotic and Surgical AI Agents}
Autonomous surgical robots and AI-assisted robotic surgery tools help enhance the precision and dexterity of surgeons. While these technologies still rely on human oversight, AI agents can predict tissue responses, reduce tremors in tool manipulation, and adjust strategies mid-operation for the best possible outcomes. Outside the operating room, robots equipped with AI-powered navigation systems can assist in rehabilitation centers, aiding patients with mobility issues and helping clinicians monitor progress more efficiently~\cite{fu2024ddn3}.

Artificial intelligencetificial intelligence agents are poised to become indispensable partners in healthcare, offering diagnostic support, patient education, and assistance with complex surgical procedures. While the road to fully realizing their potential is paved with challenges such as bias, data privacy, regulatory hurdles, and integration complexities, ongoing research and responsible innovation are making rapid progress~\cite{xu2024can}.

\subsection{Business and Industry}
Enterprise applications of AI agents span customer service chatbots, supply chain optimization, and strategic decision support. Conversational agents handle routine inquiries, freeing human operators to address more complex tasks~\cite{tur2011spoken,brown2020language}. AI-driven agents in operations management can forecast demand, manage inventory, and optimize logistics. Financial institutions employ agents for fraud detection, risk assessment, and algorithmic trading~\cite{deng2017deep}, reducing costs, improving efficiency, and enhancing overall competitiveness.

\subsubsection{Customer Service and Engagement}
AI-powered conversational agents, such as chatbots and virtual assistants are streamlining the customer experience by handling routine inquiries, resolving issues and personalizing customers. These agents enable businesses to operate 24/7, providing instant responses and freeing human agents to focus on other complex or sensitive issues. For example, having companies use AI chatbots to guide users through product selection, technical issues and managing post-purchase support\cite{satheesh2021applications}.

In addition to traditional customer service roles, AI agents are now essential to marketing campaigns. They analyze customer behavior to create targeted advertisements and personalized content. By dynamically adapting to user preferences, these agents help businesses improve conversion rates and create long-term customer loyalty, overall; this empowers businesses to scale operations without compromising quality.

\subsubsection{Supply Chain Optimization and Operation Management}
AI agents are revolutionizing supply chain management by enabling efficient forecasting, inventory control, and logistics optimization. The complex and dynamic nature of modern supply chains make them an ideal application area for AI as these systems can process vast amounts of data to identify patterns, predict disruptions, and suggest corrective actions~\cite{du2024embracing}.

Predictive analytics allow businesses to forecast demand with significant accuracy. This capability helps to minimize overproduction and stockouts, leading to significant cost savings. For instance, multinational corporations like Amazon rely on AI agents to manage their inventory globally across warehouses, dynamically adjusting stock levels based on real-time sales. \cite{christopher2022logistics}\cite{dash2019application}

When it comes to logistics, route optimization algorithms consider variables such as traffic, weather conditions and delivery deadline to find the most efficient paths for transportation. This not only reduces fuel consumption and delivery times but also contributes to sustainability goals. \cite{christopher2022logistics}\cite{dash2019application}

Warehouse automation further highlights the role of AI agents. With autonomous robots powered by machine learning and computer vision technologies while managing tasks like sorting, packing and inventory auditing. These innovations reduce human error, improve throughput and enhance overall operational efficiency, making AI agents indispensable in modern supply chain ecosystems. \cite{davenport2018ai}.

\subsubsection{Financial Decision-Making and Risk Management}

The financial sector has embraced AI agents for their ability to analyze large datasets,identify trends and make decisions with precision and speed. AI applications in finance range from fraud detection, and credit scoring to algorithmic trading and portfolio management~\cite{liu2024application}.

Fraud detection is a critical section for AI agents to excel. By analyzing patterns in transaction data, these systems can flag anomalies that might indicate fraudulent activity. For example, credit card companies use AI to detect unusual spending behavior in real time, notifying customers, and freezing accounts to prevent losses. These systems continuously learn and adapt, improving their ability to detect types of fraud as they emerge\cite{bao2022artificial,wang2025application}.

In credit risk assessment, AI agents evaluate an applicant's creditworthiness by analyzing  traditional metrics alongside alternative data sources, such as social media activity and online behavior. This holistic approach provides lenders with deeper insights, enabling them to make more informed decisions while expanding access to credit for under banked populations~\cite{zhou2025regression}.

AI agents have transformed financial trading by enabling institutions to analyze complex datasets and execute trades with precision and speed. In algorithmic trading AI systems analyze market trends, price movements and macroeconomic indicators to make split-second trading decisions that would be impossible for a human to replicate \cite{nuti2011algorithmic}.
These agents optimize strategies by backtesting historical data, identifying profitable patterns and adapting to real-time market changes. For example, reinforcement learning models are increasingly used in trading algorithms to improve performance by learning from past successes and failures \cite{10761352,feng2025ad}. Such systems enhance profitability while reducing the impact of human bias and error.

Risk management is another important area where AI agents are proving to irreplaceable. Advanced risk models powered by machine learning can assess and predict potential market risks by analyzing diverse datasets, including volatility indices, credit spreads, and geopolitical events\cite{hassan2018survey,ke2025detection}. By identifying correlations and anomalies that humans might overlook, these systems enable traders and portfolio managers to mitigate exposure to adverse market conditions. Additionally, AI agents are used to simulate stress-testing scenarios, helping financial institutions to prepare for extreme market events\cite{olorunnimbe2023deep}. This ability to provide actionable insights ensures that trading strategies remain robust and adaptive, even in highly uncertain environments.

\subsection{Education}

Recent efforts to implement AI agents in education have targeted elementary \cite{dieker2024using}, middle \cite{zha2024mentigo}, and upper level \cite{as2024intelligent} students. These efforts target diverse areas, including healthcare \cite{as2024intelligent}, language learning \cite{lan2024teachers}, and computer science \cite{wang2024generative,yunoki2023exploring}. Across these domains, use cases can be grouped into 2 main categories: elevating student engagement and reducing instructor workload. \cite{lan2024teachers} identifies 3 potential strengths of AI agents in education: team teaching, personalized learning suggestions, and data-driven feedback. In their discussion of team teaching, they argue that human teachers should provide knowledge of pedagogy, subject expertise, and emotional intelligence, while the AI agent is utilized for information processing. A main limitation is that the information provided by AI agents may still not be perfectly accurate, and neurodiverse populations are not fully represented in training data \cite{dieker2024using}. While acknowledging challenges, authors propose architecture guidelines for AI agent integration into education \cite{xu2024foundation, istrate2024ai, jiang2024ai} and are generally optimistic about future progress in the area.

\subsubsection{Student engagement}

AI agents are commonly implemented in education with a knowledge base containing course materials \cite{isaacs2024design, kumarathunga2024transforming}. One study \cite{wang2024generative} implemented AI agents as co-learners for asynchronous learning. While the students watched tutorial videos, the agents had access to their screen and mouse movements, and students interacted with the agents through text and voice message. This resulted in improved social and cognitive presence. Cognitive presence is generally defined as meaning derived through sustained reflection and discourse; social presence is perceiving one's environment to contain others. Another project, Mentigo,\cite{zha2024mentigo} adaptively responds to student interactions. The agent tracks 3 parameters: creative-problem-solving stage, student state, and selected strategy; after each dialogue round, the agent determines the state and stage, and from these values chooses the appropriate strategy.

AI agents can also help to simulate real-world scenarios. For example, in \cite{as2024intelligent}, students are studying healthcare root cause analysis, and agents representing professional roles—such as pharmacist and prescriber—provides students with the opportunity to participate in more dynamic and representative simulations than they might otherwise have access to. The authors implement a mentor agent which provides feedback to the learner based on every last 5 interactions. This mentor has access to the user's interactions with all other agents, as well as the course and assignment goals, and is thus able to guide students towards the correct answer without revealing it directly: a key educator design goal for AI agents.

\subsubsection{Educator workload}

Educators often find themselves without the resources to provide personalized attention to every student. And although full AI agent integration has not yet taken place in education, several groups have identified good practices, conducted initial experiments, and created AI agents focused on education. One group \cite{lan2024teachers} created a customized ChatGPT agent. In the system prompt, they provide the model with several steps to use while interacting with the student. For example, this group's focus was teaching order words, so their first step was "Explain what are order words?", part of their fifth step was "Confirm students’ understanding by asking them the following multiple-choice question...", and their last step was "Ask students to describe the story of Snow White in a 5-step story and use appropriate order words to tell their story.", which is the assignment evaluated by the educator. This process can be iterative, with the educator providing feedback at chosen steps, and the agent guiding students towards meeting the goals of the provided rubric while only providing educator-approved content~\cite{qu2025generative}.

Educators may also seek AI agent assistance in managing group dynamics, such as through socially shared regulation of learning (SSRL). \cite{edwards2024human} designed an agent with the task of raising group-level metacognitive awareness as part of improving SSRL, and prompted students when their contribution was not identified as group-oriented or when there were communication issues. Although the initial experiment with the group of 52 pre-service teachers was not considered successful in promoting SSRL, this work is illustrative of the future potential AI agents may have in regulating group settings.

AI agents also have the potential to assist with the development of course materials. \cite{yu2024mooc} presents one method, where the educator provides a set of slides, and the AI agents work to provide lecture notes and other resources dynamically based on student need. The study also supports a modified Massive Open Online Course (MOOC) format, termed MAIC (Massive AI Course) and employs teacher, assistant, classmate, analyzer, and manager agents as part of the core interaction infrastructure, with the additional option of custom agents as requested by the user. They also test their concept with 500 university volunteers and 2 courses, and find promising preliminary results~\cite{liu2024systematic}.

\subsection{Science and Research}
Automated laboratories and scientific discovery agents can aid in optimizing laboratory functions by designing experiments, analyzing results, and proposing new hypotheses~\cite{Kusne2023scalable}. Agents can rapidly process large datasets from genomics, proteomics, materials science, and physics, identifying patterns and suggesting novel solutions to catalyze fundamental research breakthroughs in many fields.

AI agents are able to contribute meaningfully to various aspects of the research process by making use of various cognitive and technical skills including, vast knowledge bases, self-management and evaluation, and communication. Research frameworks suggest AI agents can assist heavily in tasks involving data analysis, information retrieval, experimental design, and even the formulation of research questions and hypotheses. Multi-agent systems allow for interactions between individual agents while completing tasks, enabling seamless collaboration between human and AI researchers on all fronts~\cite{zhao2024acellular,gyager2024exocortex}.

\subsubsection{Automated Laboratories in Biology and Chemistry}
Research in Biology and Chemistry has seen the benefit of AI agent integration in automating laboratory functions. Leveraging their vast knowledge bases, AI agents can employ domain-specific knowledge when carrying out tasks such as experiment design and analysis of large datasets~\cite{shir2024towards}. This can be especially useful in materials sciences where data scarcity and comprehensive knowledge requirements can slow the research process~\cite{yu2024llm}. Furthermore, combining intelligent AI agent systems with robotics capabilities and laboratory hardware enables these systems to interact with the real world, automating repetitive lab work and minimizing risk when dealing with hazardous biological and chemical materials~\cite{song2024robot, rihm2024manager}. Multi-agent systems can be utilized to conduct virtual simulations of cellular and molecular environments alongside subsequent analysis~\cite{lu2022cot}, accelerating scientific discovery within these fields~\cite{gao2024empowering}.

\subsubsection{AI Researchers}
While the utilization of AI and AI agents for completing specific research tasks has been well studied, there has been a recent effort to expand these agents to assist in the research process more holistically. Current AI research agents are prevalent in computer science and machine learning research. With many LLMs already possessing programming knowledge and capabilities, LLM-based agents can be used to maintain codebases, interact with files, conduct testing, and perform more general research tasks (hypothesis formulation, results analysis and discussion, etc.), mostly out-of-the-box~\cite{li2024copilot}. Furthermore, Projects like the "AI Scientist" have attempted to create an agent capable of fully automating all aspects of research from generating novel research ideas to conducting experiments to publishing their findings~\cite{lu2024scientist}. Despite their remarkable potential, many of these agents still struggle with issues present in LLMS such as hallucinations and explainability~\cite{lu2024scientist,  laurent2024lab-bench}. Nevertheless, as these models improve over time and are better able to synthesize their vast amounts of knowledge, the viability of fully autonomous research agents will only grow.

\subsection{Public Services and Urban Planning}

AI agents also enhance public service delivery, from optimizing public transportation schedules to managing energy grids and water distribution systems. Urban planners use AI-driven simulations to forecast the impact of infrastructure projects, assess environmental policies, and develop sustainable city layouts. By integrating data from diverse sources, agents can balance competing objectives such as cost, efficiency, and social equity.
\subsubsection{Land Use and Urban Resource Management}
Sustainable urban growth requires efficient resource management and land utilization. In parks and gardens, sensors keep an eye on irrigation-related parameters to guarantee effective water use. Energy conservation is aided by smart street lighting systems that have wireless internet relay capabilities, CCTV, and energy-efficient LED lights. Geographic Information Systems (GIS) and geospatial technologies help with resource allocation and urban planning by offering insights into land use trends\cite{bauer2021iot}. AI helps create the best land use plans by striking a balance between environmental sustainability and development requirements. AI also facilitates the installation of green infrastructure, which reduces pollution and enhances urban sustainability~\cite{li2024exploring}. Examples of this include parks and green roofs. which help mitigate pollution and improve urban sustainability
\subsubsection{Collaborative Decision-making AI Agents in Public Administration}
In public administration, AI and participative sensing technologies greatly improve decision-making. By enabling people to report events or incidents in the city, participatory sensing enhances public administration's responsiveness and efficacy by sharing the reports with other users\cite{bauer2021iot}. AI-driven urban planning offers data-driven insights that assist planners in making well-informed choices to enhance air quality and efficiently control urban growth. Predictive models help public administration improve public safety and allocate resources effectively by forecasting high crime risk locations\cite{kouziokas2017application}. Data-driven policymaking is supported by AI and GIS technologies, which offer insightful information for public administration and urban planning\cite{zheng2024urban}. Policymakers can use these insights to inform policies that minimize pollution and advance sustainable development.

\subsubsection{Transport routing optimization AI agents}
AI technologies are essential for optimizing traffic flow and minimizing congestion in urban transport routes. Parking sensors save needless travel and pollution by guiding drivers and detecting available parking places. In order to help with effective traffic management, traffic intensity monitoring devices evaluate vehicle speeds, road occupancy, and traffic volumes. AI optimizes effective transportation flows, forecasts traffic, and examines transportation patterns. Through the support of Transit-Oriented Development (TOD), artificial intelligence (AI) contributes to the development of transit networks that provide high-capacity, safe, and effective modes of transportation while lowering emissions and traffic congestion\cite{zheng2024urban}. Furthermore, AI applications in transportation safety improve public safety and optimize transportation routes, for example, by forecasting high-crime risk zones\cite{kouziokas2017application}.
\subsection{Entertainment and Creativity}
In the 21st century, people's entertainment has dramatically shifted from traditional offline mediums to digital platforms, like social media platforms, video/music streaming services, and video games across different platforms and devices. This evolution has further been enhanced by recommendation algorithms specifically designed to personalize user experiences. Nowadays, the integration of AI-agents into the entertainment industry is becoming more popular. It's a new trend that people start using AI-agents to create a better experience for the users.
In the entertainment industry, AI-agents not only generate personalized content recommendations, but also drive non-player characters in games. They can even assist in creative processes such as music composition, artwork generation, and scriptwriting as well. These creative agents leverage cutting-edge large language models and multimodal generation, along with traditional neural networks and optical character recognition techniques to support and produce results which are tailored to users' interests and needs~\cite{schatten_ai_2024}. Such personalized tools opens new paths for innovative approaches to artwork creation and interactive storytelling and significantly boosts players' engagements and entertainments in games.

\subsubsection{Visual Design and Storytelling}
Artists and authors have long benefited from artificial intelligence. Tools such as text generation from ChatGPT and image creation from MidJourney and Stable Diffusion have reshaped the production of high-quality artworks and stories by aiding in the visualization of ideas, and reduction of repetitive tasks~\cite{ding_designgpt_2023} This evolution of workflow is further enhanced by the integration of AI-agents, which has the ability to integrate with multiple tools simultaneously to deliver design inspirations via both visual and textual mediums~\cite{ding_designgpt_2023}. Additionally, there is Mimisbrunnur, which employs a mixed-initiative system to assist the writers in crafting compelling, high-quality interactive stories~\cite{stefnisson_mimisbrunnur_2018}.

\subsubsection{Video games}
Video games stand as a dominating form of entertainment in modern society. Their origin can track back to the 1960s. In order to create a more living environment and interactive experience, the concept of computer-controlled non-player characters (NPCs), inspired by Dungeons and Dragons, was introduced in the late 1970s. In the early years, the decision-making process of NPCs relied heavily on algorithms like Finite State Machines, as seen with Pac-Man's ghosts, and tree-based methods like decision trees and behavior trees~\cite{uludagli_non-player_2023}. Today, people have seen the potential power of AI-agents and therefore shifted focus toward utilizing AI-agents to create more unique and revolutionary gaming experiences for players. Projects like the Scalable, Instructable, Multiworld Agent (SIMA) aim to develop a generalized AI-agent that can learn and engage with open-world games using a combination of on-screen visuals, text information as inputs, and potential user language directionss~\cite{team_scaling_2024}. Another study explores AI-agents in the game "Passcode," where the AI-agent plays the role of the "giver", providing clues for the "guesser", which is the player, to guess a randomly chose word at the beginning of the game, demonstrating AI-agent's ability to understand and handle interactive tasks~\cite{gero_mental_2020}.

However, while there's huge potential in integrating AI-agents into the games, such a process still faces significant challenges. Currently, the behavior of AI-agent, particularly those driven by Large Language Models and neural networks, still struggles to reach the skill level of experienced human players. These AI-agents can be easily identified due to their unnatural actions and movements~\cite{g_michael_youngblood_agent-based_nodate}. In addition, although they can understand simple commands like "move forward" and "stop", they struggle to understand more complicated tasks~\cite{team_scaling_2024}. Nevertheless, ongoing research still promises a rapid evolution in AI-agent performance within the gaming industry.

\subsection{Societal Impact and Considerations}
The applications of AI agents show significant promise and, at the same time, the social aspects cannot be ignored. It is crucial to guarantee the fairness, privacy, and inclusion of individuals in the design and decision making processes of AI technologies~\cite{hagendorff2020ethics,jobin2019global}. The public, industry leaders, and policymakers must work together to create and establish ethical principles, rules, and regulations that should be followed in the development and use of AI technologies.

The AI revolution has the potential to surpass the previous industrial and digital revolutions in scale and impact. While the Industrial Revolution mechanized manual tasks and the Digital Revolution automated routine mental tasks, the AI revolution has the potential to replicate and improve nearly all human cognitive functions~\cite{makridakis2017forthcoming}. Apparently, AI agents have the most promising role in accelerating the AI revolution because they can invent new products and processes, increase efficiency, and apply them to various sectors.

However, with these promising opportunities, the AI revolution, driven by the rise of advanced AI agents, would bring significant challenges that governments and societies must address. AI agents, with their ability to automate not just routine but also complex cognitive tasks, can accelerate job displacement across industries. Their integration could disproportionately impact workers whose skills are rendered obsolete, amplifying skill mismatches and economic disruptions, particularly in regions heavily reliant on automation-vulnerable sectors~\cite{makridakis2017forthcoming}. Furthermore, as AI agents become integral in decision-making processes, the economic gains they generate are likely to concentrate among those who develop and control these systems, exacerbating wealth inequality both within and across nations.

Unlike traditional large language models, AI agents operate autonomously and adapt over time, which makes accountability far more complex. Pin down who or what is responsible for an agent’s decisions or actions becomes increasingly difficult as its behavior evolves. Because these agents can interact directly with people, other systems, and even their surroundings, their influence tends to be much broader and more systemic. This amplifies both their potential impact and the risk of unintended outcomes.

To address the social impacts brought by AI Agents, it is important to develop standardized evaluation methods that ensure consistency and reliability in assessing potential harms and benefits. Transparency in AI development, data usage, and deployment must be established to build trust. Inclusive policies and equitable resource distribution are crucial to mitigating harms and reducing disparities. Furthermore, promoting ethical AI development requires attention to cultural and contextual differences, ensuring that AI systems respect diverse values~\cite{solaiman2023evaluating}.

In summary, AI agents are transforming a wide range of domains, improving efficiency, accessibility, and personalization. As these agents mature, they hold the potential to address grand challenges in healthcare, education, sustainability, and beyond, provided that developers and stakeholders maintain a steadfast commitment to ethical and human-centric design. Governments should also play a critical role in establishing clear regulatory frameworks, promoting equitable access to AI technologies, and ensuring accountability to safeguard societal interests while fostering innovation.

\section{AI Agent Design}
The development of next-generation AI agents:
\begin{figure}[h]
    \centering
    \includegraphics[width=1\textwidth]{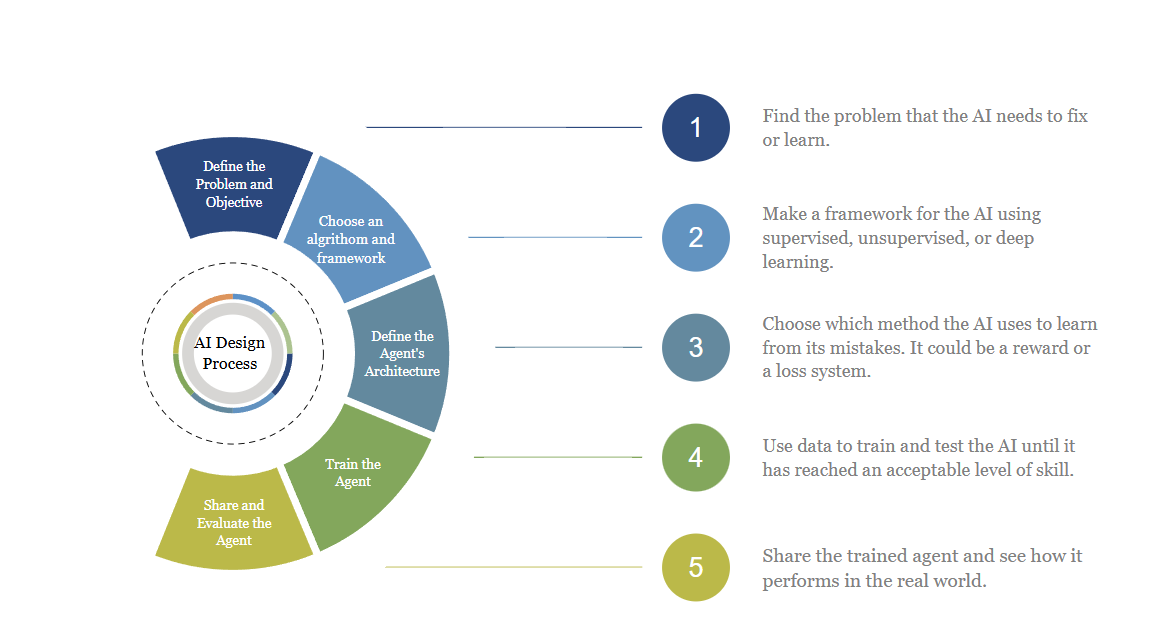}
\end{figure}

\subsection{Cognitive-Inspired Architectures}
Hybrid models that combine symbolic reasoning with neural networks are heavily inspired by human cognitive processes, where both structured reasoning and adaptive learning coexist. By leveraging the compositionality and hierarchical structure inherent in symbolic representations, these architectures not only enhance interpretability but also improve generalization capabilities across diverse tasks~\cite{lake2017building,parker2024mapping}. 

These models excel in domains requiring precise reasoning, such as scientific discovery, automated theorem proving, and natural language understanding, where traditional deep learning approaches struggle with ambiguity or long-range dependencies~\cite{marcus2020nextdecade}. The integration of neural networks enables these hybrid systems to process unstructured data like images, audio, and text, while symbolic components offer a framework for logic, abstraction, and transfer learning.

Recent advances in hybrid architectures have shown promise in areas such as program synthesis, where symbolic reasoning is used for code generation and verification, and robotics, where symbolic planning enhances decision-making in dynamic environments. Future research is expected to focus on scalable implementations of these systems and developing methods to dynamically balance symbolic and neural components for optimal performance in real-world applications ~\cite{parker2024mapping}.

\subsection{Hierarchical and Modular Approaches}

Decomposing complex tasks into manageable subtasks and leveraging modular architectures fosters scalability, reusability, and system stability. Hierarchical frameworks enable AI agents to break down high-level goals into smaller, more tractable components, streamlining task execution and improving performance. Modular architectures, on the other hand, allow for the development of specialized submodules, each dedicated to a specific functionality, thereby promoting adaptability and efficient resource allocation.

These approaches also enhance interpretability by allowing researchers and practitioners to understand the role and behavior of individual modules, which is particularly critical in safety-critical domains like autonomous vehicles, healthcare, and robotics. Additionally, modular systems simplify failure diagnosis, as issues can often be traced back to specific components rather than requiring analysis of the entire system.

Recent advances have demonstrated the utility of hierarchical and modular designs in reinforcement learning, natural language processing, and robotics. For instance, hierarchical reinforcement learning algorithms decompose decision-making processes into macro-actions, enabling agents to solve long-horizon tasks more effectively. Similarly, modular neural networks in natural language processing have been employed to handle tasks like machine translation, sentiment analysis, and summarization within a unified framework.

Moving forward, future research should focus on enhancing the dynamic interplay between hierarchical control layers and modular subsystems, enabling AI agents to seamlessly adapt to new and unforeseen challenges. Efforts to integrate these approaches with techniques such as meta-learning and transfer learning hold promise for creating robust, scalable, and generalizable AI agents capable of operating in complex, real-world environments ~\cite{huang2022language,li2023learning}.

\section{Discussion}


\subsection{Discussion and a Step-by-Step Guide}

For newcomers, AI agent research can seem overwhelming due to its interdisciplinary nature. This section provides a structured path to help researchers gain foundational knowledge and practical experience. Given its breadth and interdisciplinary nature, newcomers often struggle to find a structured learning path. This guide provides an approach to help researchers navigate challenges, build foundational knowledge, and develop impactful projects.\\

\textbf{Step 1: Build a Strong Theoretical Foundation}\\ 

Developing a solid theoretical foundation is the first step. Understanding how AI agents perceive, make decisions, and act is essential before implementation. Topics such as reinforcement learning, multi-agent coordination, planning, and decision theory provide a necessary framework.

Foundational works such as \textit{Multi-Agent Systems: A Survey} and \textit{Reinforcement Learning: An Introduction}~\cite{sutton2018reinforcement} help researchers grasp core AI agent architectures and methodologies. Online courses and research blogs, such as \textit{Lilian Weng’s Blog}~\cite{weng2023agent}, provide accessible explanations of advanced topics like memory-augmented neural networks and tool-augmented learning.

Beyond theory, implementing simple reinforcement learning algorithms in controlled environments is crucial. OpenAI Gym and similar frameworks allow hands-on experimentation, helping researchers develop an intuitive understanding of AI agent behavior. Starting with basic agents for navigation, decision-making, and task automation reinforces theoretical knowledge through direct application~\cite{ding2024enhance}.\\

\textbf{Step 2: Focus on Measurable Projects}\\ 

AI agent research requires clear success metrics. Unlike traditional supervised learning, AI agents operate in dynamic environments where evaluation methods differ. Researchers should begin with structured projects that allow controlled experimentation and measurable outcomes.

Single-agent reinforcement learning tasks, such as autonomous navigation and game-playing, provide an excellent starting point. Evaluating performance through reward signals, convergence rates, and task completion times ensures systematic progress tracking. For those interested in multi-agent research, simulation platforms such as CARLA~\cite{github:carla} and PettingZoo~\cite{github:pettingzoo} offer controlled settings to study collaborative and competitive agent behaviors.

Defining evaluation metrics early ensures research remains focused and reproducible. Whether optimizing decision policies, improving sample efficiency, or enhancing interpretability, measurable goals help refine research directions and improve experimental reproducibility.\\

\textbf{Step 3: Gain Practical Experience with Tools and Iterative Learning}\\

Hands-on experience with AI agent frameworks is essential for moving beyond theoretical learning. Simulation environments such as RoboCup Soccer Simulation~\cite{github:robocup} and CARLA~\cite{github:carla} allow structured experimentation without extensive development, enabling researchers to focus on agent decision-making, coordination, and adaptability~\cite{chen2025amazon}.

AI agents learn through repeated interactions rather than fixed datasets. Iterative learning techniques, such as reinforcement learning with self-reflection and hierarchical RL~\cite{weng2023agent}, help refine agent behavior. Adjusting hyperparameters, modifying architectures, and fine-tuning reward functions improve learning efficiency and stability. 

Engagement with open-source AI communities accelerates learning. Platforms like GitHub host repositories with baseline models and experimental frameworks, allowing new researchers to explore existing implementations and contribute to ongoing projects. Reproducing research paper results is another valuable exercise, helping researchers understand the nuances of implementation and evaluation~\cite{hao2024artificial}.\\

\textbf{Step 4. Leverage Open-Source Communities and Reproducibility.}\\

Engagement with open-source projects accelerates learning and fosters collaboration. Sharing project code and insights within the community not only enhances reproducibility but also allows researchers to build on each other's work. Platforms like GitHub host numerous repositories for AI agent research, providing access to baseline models and implementations that can serve as starting points for new projects. Projects like CARLA and PettingZoo provide a variety of tools including access to simulation environments, assets, models, and step-by-step instructions, enabling a straightforward process for replicating experiments and streamlining iteration upon existing work~\cite{github:carla, github:pettingzoo}. Projects like RoboCup Soccer Simulation leverage open-source competitions to drive community engagement and accelerate development of cutting-edge AI agent systems. By actively contributing to these communities, researchers can stay informed about emerging trends and gain feedback on their work~\cite{yang2024hades}.\\

\textbf{Step 5. Address Broader Research Gaps.}\\

The novelty of AI agents means that many areas remain underexplored. Researchers should identify gaps in the literature, such as the integration of advanced planning algorithms with tool-augmented LLMs or the use of hybrid symbolic-subsymbolic approaches for better interpretability~\cite{marblestone2016toward, lake2017building}. Exploring these areas can lead to groundbreaking contributions and help establish best practices for the field.

\subsection{Challenges and Limitations}
Despite substantial progress, significant hurdles remain that limit their potential~\cite{zhang2024optimizationapplicationcloudbaseddeep}. This section examines critical areas of concern, including safety, interpretability, ethics, generalization and acability, and highlights the need for interdisciplinary efforts to address these issues:

\subsubsection{Safety and Robustness}
Among these, ensuring safety and robustness in dynamic environments is a persistent hurdle. AI agents often struggle with adapting changes in the environment, particularly when exposed to scenarios or data that differ significantly from their training. These vulnerabilities are further compounded by the potential for adversarial attacks, where minor input perturbations can drastically alter agent behavior.  resilience to variability and ensure reliable performance even under extreme conditions
Agents often exhibit sensitivity to distributional shifts, adversarial perturbations, and environment changes~\cite{amodei2016concrete}. Ensuring safe operation in open-ended and uncertain settings requires improved robustness techniques, formal verification, and worst-case guarantees under extreme conditions.

\subsubsection{Explainability and Interpretability}
As agents grow more complex, their decision-making processes become opaque~\cite{doshi2017interpretability}. This lack of interpretability poses significant challenges in fostering user trust, debugging errors, and complying with regulatory requirements, especially in domains such as finance and medicine. Tools for model introspection, attention visualization, causal attribution, and symbolic distillation have been proposed as potential solutions~\cite{yi2018enhance}. These approaches aim to provide clear insight into agent decision-making, enabling stakeholders to better assess and trust system's outputs. However, current methods remain limited and require further refinement to align with practical needs of end-users and regulatory frameworks.

\subsubsection{Ethical and Social Considerations}
The deployment of AI agents in sensitive domains introduces ethical and social challenges. Issues such as bias, fairness, privacy, and accountability are pressing concerns as agents often inherit biases present from their training data, perpetuating social inequalities  ~\cite{jobin2019global,hagendorff2020ethics}. For example, the \textit{MDPI review on AI agent challenges} discusses these concerns in detail, emphasizing the need for transparent evaluation methods and robust regulatory frameworks. Aligning agents with human values, establishing standards for responsible AI, and developing frameworks for ethical oversight will be crucial as agents increasingly connect and interact with social systems. Without these safeguards, the widespread deployment of AI agents could exacerbate existing disparities and raise concerns about their impact on human autonomy and rights~\cite{su2024large,jin2024scam}.



\subsubsection{Generalization and Transfer}
Many agents struggle with out-of-distribution generalization, requiring extensive retraining for new tasks or domains, often requiring extensive training to operate effectively in different environments. This lack of transferability undermines their utility in real-world applications, where agents must adapt quickly to novel scenarios~\cite{zhao2021towards}. For example, an agent trained in a simulated environment may fail to perform effectively in a real-deployment due to subtle but critical differences between two contexts. Advancing methods for domain adaptation, transfer learning, and task-agnostic representations will be essential for creating versatile and agile agents capable of functioning in diverse and unpredictable conditions~\cite{liu2024seamcarver,chen2022relax}.

\subsubsection{Scalability and Resource Efficiency}
Training and deploying state-of-the-art models can be computationally and energy-intensive which represents a barrier to scalability and accessibility. Modern AI agents often require extensive resources, limiting their deployment to organizations with substantial computational infrastructure~\cite{strubell2019energy}. Additionally, the environmental impact of these systems raises concerns about sustainability. Research on model compression, efficient architectures, and distributed learning paradigms aims to reduce resource footprints without sacrificing performance. These advancements are necessary to make AI agents more accessible to smaller organizations and environmentally sustainable for widespread adoption~\cite{xiang2023tkil,zhang2024federated,YANG2024102380}.

\subsection{Future Directions and Emerging Opportunities}
Despite current challenges in the field of AI agents, several promising frontiers offer opportunities for future advancements in this area. Therefore, this section explores some of these potential applications where AI agents could support and enhance research and production~\cite{Yang2024}:

\subsubsection{Neuroscience-Inspired Mechanisms}
Recent studies have applied principles from neuroscience into deep learning, particularly focusing on how neurons in the brain utilize cost functions to adapt to diverse contexts by changing their properties. This integration has demonstrated utility in enhancing the performance of deep learning models ~\cite{marblestone2016toward}. Consequently, this direction worth future attention to dive deeper into the adoption of additional neuroscience paradigms, including predictive coding, dendritic computations, and synaptic plasticity, into AI agents. Such integration could potentially yield more stable, interpretable, and efficient learning mechanisms~\cite{zhao2024towards}.

\subsubsection{Interactive and Continual Learning}
Traditional machine learning approaches are mainly designed for static environments, such as training models to classify predefined categories or to play specific games. A significant drawback of these methods is catastrophic forgetting, where models lose previously learned concepts when exposed to new data or features~\cite{chen2025association}. In response, research has examined the concept of continual learning, an architecture that involves learning from an endless stream of data while retaining previous knowledge ~\cite{finn2017model,you2022incremental}. This approach has proven useful for machine learning models for keeping a long-term memory on learned features, suggesting that it could enable AI agents to refine their knowledge base through iterative feedback, human demonstrations, and structured curricula, and consequentially making them more robust and versatile ~\cite{lin2025slam2,de2019continual,10.1145/3627673.3679997}.

\subsubsection{Hybrid Symbolic-Subsymbolic Models}
Symbolic and subsymbolic models represent the two primary types of AI models, categorized by their background features like reasoning and knowledge storage mechanisms. Both types has demonstrated significant success in performing specific tasks~\cite{dan2024multiple,lin2024dpl,zhuang2025gradient}. Given these strengths, it is worthwhile to explore the integration of these methodologies and leveraging the advantages of both. We believe such an approach of blending the transparency and structure of symbolic reasoning with the pattern recognition capabilities of deep neural networks holds the potential for robust generalization, interpretability, and efficiency~\cite{lake2017building,marcus2020nextdecade}.

\subsubsection{Multi-Agent Governance and Coordination}
The architecture of AI agents inherently involves the collaboration of multiple agents, each possessing unique features and advantages. As the scale of an AI agent system grows, the number of participants would increase~\cite{li2024political}. Thus, the governance and coordination of these agents — which includes task allocation, negotiation protocols, and data race management — are crucial to the system's performance. Studies have indicated that by implementing suitable protocols within multi-agent deep learning models, system performance can be enhanced~\cite{yang2018mean,zhong2025enhancing,wang2024pargo}. Therefore, it is essential to conduct future research focused on these coordination mechanisms to advance the performance of AI agents~\cite{huo2025ctpatchtstchanneltimepatchtimeseries,jiwadflockjs}.

\section{Conclusion}
AI agents have transformed from specialized, rule-bound systems to increasingly integrated, autonomous entities that perceive, reason, act, and collaborate. This review surveyed the historical evolution, core architectural components, and emerging paradigms that define contemporary AI agents. We discussed breakthroughs in reinforcement learning, large language models, hierarchical planning, and embodied intelligence. Yet, critical challenges persist: improving safety, interpretability, and ethical stewardship, as well as achieving robust generalization and resource efficiency~\cite{zhu2024podb,chen2022explain,zhao2024towards}.

The path forward demands interdisciplinary engagement. Insights from cognitive science, neuroscience, sociology, economics, and ethics will inform next-generation agents. By prioritizing human values, transparency, and long-term adaptability, we can usher in an era where AI agents serve as trustworthy partners in scientific research, industrial automation, healthcare, education, and beyond. With sustained collaboration and careful innovation, the future of AI agents holds the promise of more capable, responsible, and beneficial autonomous intelligence~\cite{elmachtoub2023estimate,10.1145/3627673.3679071}.


\section*{Competing Interests}
The author declares no competing interests.

\appendix
\section{Appendix}

The first 4 queries were used to query Google Scholar and the last query was used to query Papers with Code:\\

\noindent \textbf{Business}: \textit{("AI Agent" OR "ML Agent" OR "Deep Learning Agent" OR "Autonomous Agent" OR "Automated AI" OR "Generative AI Agent" OR "Reinforcement Learning Agent" OR "Intelligent Agent" OR "Multi-Agent System")
AND
("Business" OR "Industry" OR "Industrial Applications" OR "Manufacturing" OR "Production" OR "Supply Chain" OR "Logistics" OR "Automation" OR "Process Optimization" OR "Predictive Maintenance" OR "Operations Management" OR "Enterprise AI" OR "Decision-Making in Industry" OR "Industrial AI")”
}\\

\noindent \textbf{Education}: \textit{(‘AI Agent’) AND ('Education') AND ('Learning' OR 'Teaching') AND ('Classroom' OR 'Critical Thinking' OR 'Curriculum'  OR 'Synchronous') AND ('Resource')}\\

\noindent \textbf{Science}: \textit{("AI Agent" OR "Agent" OR "Autonomous Agent" OR "ML Agent" OR "Multi-Agent System" OR "Reinforcement Learning Agent" OR "LLM-Based Agent") AND ("AI Scientist" OR "AI Researcher" OR "Autonomous Researcher" OR "Autonomous Laboratory" OR "Scientific Discovery Agent")}\\

\noindent \textbf{Entertainment}: \textit{("AI Agent" OR “AI-Assisted”) AND (“Entertainment” OR “Design” OR “Game” OR “Storytelling”)}\\

\noindent \textbf{Social Impact}: \textit{(“AI Agent” OR “Generative AI”) AND (“Social Impact” OR “Society” OR “Social Implications”)}\\

\noindent \textbf{Papers with Code}: \textit{"AI Agent"}

\bibliography{reference}
\end{document}